\documentclass{elsart}

\def\etal{{et al.~}}
\def\eg{{e.g.,}}
\def\ie{{i.e.,~}}
\def\kms{~{\rm km~s^{-1}}}
\def\cm3{~{\rm cm^{-3}}}
\def\Myr{~{\rm M}_{\odot}~yr^{-1}}
\def\Msun{~{\rm M}_{\odot}}
\def\Zsun{~{\rm Z}_{\odot}}
\def\lsim{\mathrel{  
        \raise0.3ex\hbox{$<$}\kern-0.75em{\lower0.65ex\hbox{$\sim$}}}}
\def\gsim{\mathrel{
        \raise0.3ex\hbox{$>$}\kern-0.75em{\lower0.65ex\hbox{$\sim$}}}}

\usepackage{natbib}
\usepackage{graphicx}
\usepackage{amssymb}

\journal{New Astronomy}

\begin{document}

\begin{frontmatter}

\title{Feedback from Multiple Supernova Explosions inside a Wind-Blown Bubble} 

\author{Hyunjin Cho \&}
\ead{lyrae@pusan.ac.kr}
\author{Hyesung Kang\corauthref{cor} }
\address{Department of Earth Sciences, Pusan National University, 
Pusan 609-735, Korea}
\ead{kang@uju.es.pusan.ac.kr}
\corauth[cor]{Corresponding author.}

\begin{abstract}

We study the evolution of multiple supernova (SN) explosions inside a pre-exiting
cavity blown by winds from massive progenitor stars.
Hydrodynamic simulations in one-dimensional spherical geometry, including
radiative cooling and thermal conduction, are carried out 
to follow first the development of the wind-blown bubble during the main 
sequence and then the evolution of the SN-driven bubble. 
We find the size and mass of the SN-driven bubble shell depend on the 
structure of the pre-existing wind bubble as well as the SN explosion energy 
$E_{SN}$ ($= N_{SN} 10^{51}{\rm ergs}$).
The hot cavity inside the bubble is 2-3 times bigger in volume and hotter  
than that of a bubble created by SNe exploded in a uniform interstellar medium (ISM).
For an association with 10 massive stars in the average ISM, 
the SN-driven shell has an outer radius of 
$R_{ss} \approx (85 {\rm pc}) N_{SN}^{0.1}$ and 
a mass of $M_{ss} \approx (10^{4.8} \Msun)  N_{SN}^{0.3}$
at $10^6$ years after the explosion.
By that time most of the explosion energy is lost via radiative cooling, 
while $\lsim$ 10\% remains as kinetic energy and $\sim $10\% as 
thermal energy.
We also calculate the total integrated spectrum of diffuse radiation emitted 
by the shock-heated gas of the SN bubble.
Total number of H Lymann-limit photons scales roughly 
as $\Phi_{13.6} \approx 10^{61} N_{SN}$
and those photons carry away 20 - 55 \% of the explosion energy.
For the models with 0.1 solar metalicity, the radiative energy
loss is smaller and the fraction of non-ionizing photons is larger,
compared to those with solar metalicity.   
We conclude the photoionization/heating by diffuse radiation is the most 
dominant form of feedback from SN explosions into the surrounding medium.

\end{abstract}


\end{frontmatter}

\section{Introduction}

Massive stars inject a significant amount of energy into the surrounding medium 
in various forms: photoionization/heating by stellar radiation and
mechanical energy through stellar winds and supernova (SN) explosions,
some of which later transforms into gas thermal energy and radiation.
For example, OB stars deposit stellar radiation energy into the
interstellar medium (ISM) with the luminosity of $10^4 \lsim L_*/L_{\odot} 
\lsim 10^6$ and
the H Lymann-limit photon luminosity ($h\nu>13.6$eV) of 
$ 10^{47} {\rm s^{-1}} \lsim Q_{\rm H} \lsim 10^{50} {\rm s^{-1}}$ \citep{smi06}.
So the radiative feedback of an early type star during the main sequence
($\tau_{\rm MS} \sim 10^6$ yrs) may amount up to $\sim 10^{51-52} $ergs,
which is comparable to typical SN explosion energy.

These stars also inject mechanical energy into the ISM via stellar winds during the
main sequence (MS) and the red supergiant (RSG) or luminous blue variable (LBV) phase, 
and also Wolf-Rayet stage in the case of very massive stars with $M_*>25\Msun$
\citep{chu04}.
The MS winds have a typical wind speed of $v_w \sim 1000-2500 \kms$ and a mass
loss rate of $\dot M_w \sim 10^{-7}-10^{-6} \Myr$, while
the RSG winds have much lower speed, $v_w\sim 10-50 \kms$,
but higher mass loss rate, $ \dot M \sim 10^{-5} -10^{-4} \Myr$
\citep{chu04,dwa05,smi06}.
Owing to much longer duration the MS wind generates a bigger bubble,
while the RSG wind produces a much smaller but denser shell inside 
the MS bubble created earlier.
The mechanical energy luminosity of the MS winds ranges
$100 \lsim L_{\rm wind}/L_{\odot} \lsim 3000$ \citep{smi06}.

Massive stars with mass $M \gsim 8 M_{\odot}$ end their lives as core-collapse
SNe and deposit typically $E_{SN}\sim 10^{51}$ ergs in
the form of kinetic energy of the ejecta.
Initially the SN ejecta expands freely into the wind-blown cavity 
and later interacts with the shell produced previously by the winds, 
generating various shocks, \eg ~transmitted and reflected shocks.  
Most of the explosion energy is thermalized via such shocks.
After the shock-heated gas cools adiabatically down to
$10^6$ K or after the forward ejecta shock catches up with the wind bubble shell,
the supernova remnant (SNR) begins to lose a significant fraction 
of thermal energy via radiative cooling.
The final energy budget deposited into the surrounding medium
in the forms of thermal, kinetic, and radiative
energy depends on the detailed history of radiative cooling and
heat transport by thermal conduction, which in turn 
depends on the size and structure of the preexisting wind bubble 
as well as on the ambient ISM.
For example, if a SN explodes in a dense environment, the explosion energy 
would be lost mostly by radiative cooling and the impact to the ISM should be minimal.
According to \citet{ten90b}, 
in the case of SN explosions inside wind-blown bubbles in the typical ISM
with $n_{H,0}= 1\cm3$, the remaining kinetic energy fraction is 
$ E_{k} \sim 0.3 E_{SN}$ at the end of SNR evolution,
while the remaining thermal energy fraction,
$E_{th} \sim 0.2-0.4 E_{SN}$.

Feedback from massive stars should play a significant role in the formation and
evolution of galaxies. 
Especially, feedback from the first-generation SNe affects the formation and
evolution of proto-galaxies and the large scale structure in the early Universe
\citep [\eg][and references therein]{kay02,sca06}.
Although SN feedback is an important ingredient for galaxy formation, 
it occurs typically on length and time scales much
smaller than grid spacings and time steps of numerical simulations. 
So such subgrid physics is often treated through 
phenomenological prescriptions in numerical simulations for galaxy formation
and large scale structure formation \citep{cen06}.
Either `thermal' feedback \citep [\eg][]{kat92} or `kinetic' 
feedback \citep [\eg][]{nav93} with a feedback efficiency parameter 
$0<\epsilon<1$,
which is the fraction of SN explosion energy deposited in the 
form thermal or kinetic energy, were tried with varying degrees of success. 
For a thermal feedback model, the radiative cooling must be turned off 
artificially for a period of time
in order to prevent the dense gas from re-radiating the deposited energy immediately.
\citet{kay02} showed that both thermal and kinetic feedback models can 
alleviate the so-called ``catastrophic cooling'' problem of galaxy 
formation.
Thus in order to construct a physical subgrid prescription for SN feedback 
in numerical studies of galaxy formation it is necessary to estimate
quantitatively the feedback efficiency parameters,
$\epsilon_{th}$, $\epsilon_{kin}$, and $\epsilon_{rad}$ for
thermal, kinetic and radiation energy, respectively. 

Massive stars often form in an association, so multiple explosions 
of core-collapse SNe can create a large superbubble
\citep{tom86,ten90a,smi01}. 
Multiple SN explosions in the proto-galactic environment, in particular,
may have played a significant role in the formation of halo globular clusters.
In a self-enrichment model of metal-poor globular clusters, 
proto-globular cluster clouds are assumed to be enriched by multiple SN explosions 
from massive first-generation stars contained within the clouds
and later second-generation stars form in the cooled shell driven 
by the SN explosions \citep [\eg][and references therein]{par99}.
However, some have argued that the binding energy of a proto-globular cluster 
cloud may not be large enough to survive a few SNe, 
and that the homogeneity of metalicity within a globular cluster 
indicates that pre-enrichment must have occurred within larger systems 
\citep{sea78}.
According to numerical studies of early structure formation,
first objects might form at a redshift of $\sim 30$ with
a typical mass of $10^6\Msun$ and first-generation stars 
enrich the protogalactic clouds before they merge into larger clouds of
$10^{12}\Msun$ via gravitational clustering \citep{bro04}.
This favors the model that combines the pre-enrichment by first
stars before the formation of protogalactic halos and the
self-enrichment by next generation SNe within proto-globular cluster clouds. 
In this scenario the stability of the cooled shell and further star 
formation critically depends on how efficiently the SN energy is radiated 
away during the SNR evolution.

Thus in the present work we attempt to estimate feedback 
effects of core-collapse SNe, including multiple detonation, 
through numerical hydrodynamic simulations. 
We first calculate the creation of a bubble by stellar
winds from multiple massive stars ($N_{star}=1-50$) during the precedent MS stage.
The end results of the MS wind stage are adopted as the initial states of 
the SN explosion simulations.
We then make quantitative estimates of the fractions of explosion energy 
transferred into the ISM in different forms, 
\ie thermal, kinetic, and radiation energy.
Diffuse radiation emitted by the shock-heated gas, especially, 
ionizing radiation for hydrogen and helium atoms 
are calculated.

In \S 2, the basic features of our numerical simulations, 
including radiative cooling and thermal conduction, are described.
We then present the numerical results of wind bubble models in \S 3, 
and multiple supernova models in \S 4.
A brief summary is given in \S 5.

\section{Numerical Method}

\subsection{Hydrodynamic Equations}

We solve the hydrodynamic conservation equations of the ideal gas in 
one-dimensional spherical coordinate system, 
including radiative cooling and thermal conduction :
\begin{equation}
{\partial \rho \over \partial t}  +  {1 \over r^2} {\partial\over \partial r} 
(r^2 \rho u)
= 0,
\end{equation}
\begin{equation}
{\partial (\rho u) \over \partial t}  +  {1 \over r^2} {\partial\over \partial r} (r^2 \rho u^2) 
= - {\partial P \over \partial r},
\label{mocon}
\end{equation}

\begin{equation}
{\partial (\rho e) \over \partial t} + {1 \over r^2} {\partial \over \partial r}
\left[ r^2 (\rho e u + P u) \right] =
- \Lambda -  {1 \over r^2} {\partial \over \partial r}(r^2S),
\end{equation}

where $u$ is the radial velocity, $e = {P}/{[\rho(\gamma-1)]}+ u^2/2$
is the total energy of the gas per unit mass, 
$\Lambda$ is the energy loss rate per unit volume,
and $S$ is the radial heat flux.
A ratio of specific heats is assumed to be $\gamma = 5/3$.
The gas temperature is determined by $T= \mu m_H P/(\rho k_B) $, 
where $k_B$ is Boltzmann's constant, $m_H$ is the mass of hydrogen nucleus 
and $\mu$ is the mean molecular weight.

We use a grid-based, Eulerian hydrodynamics code in the
spherical symmetric geometry, based on the
Total Variation Diminishing (TVD) scheme \citep{ryu93}.
The TVD scheme solves a hyperbolic system of gasdynamical conservation
equations with a second-order accuracy.
It has been widely used for a variety of astrophysical problems
\citep[\eg][]{bae03}.

\subsection{Radiative Cooling and Thermal Conduction}

We define the cooling rate as
\begin{equation}
{\Lambda(T, n_H) = L(T)n_H^2}
\end{equation}
where $n_H$ is the number density of hydrogen nuclei.
For the cooling rate coefficient, $L(T)$, we adopt 
the non-equilibrium radiative cooling rate for an 
optically thin gas that is calculated by following the 
non-equilibrium collisional ionization of the gas 
cooling from $10^{8.5}$K to $10^4$ K under
the isobaric condition \citep{sut93}.
We consider cases with the solar metalicity ($\Zsun$) and with
1/10 of the solar metalicity ($0.1\Zsun$). 

Although we calculate diffuse radiation emitted by the shock-heated gas,
we do not explicitly follow the radiative transfer of diffuse radiation 
and photoionization of the gas.
Instead we assume $L(T)=0$ for $T \le 10^4$K and set the minimum 
temperature at $T_{min}=10^4$K.
This is in effect equivalent to assuming thermal equilibrium at $10^4$K
between radiative cooling and photoionization heating.

After completing the hydrodynamic part, updating hydrodynamical quantities,
radiative cooling is applied to the thermal energy per unit volume, $E$, 
as a separate step \citep{bae03}:
\begin{equation}
{E^{n+1}=E^n \cdot \exp(- {\Delta t \over {t_{cool}}})}
\end{equation}
where $t_{cool}= E / \Lambda$ is the cooling time scale
and $\Delta t$ is the integration time step. 
This is designed to follow approximately the thermal history of the gas 
that cools radiatively and approaches the specified minimum temperature 
in a time scale much shorter than the dynamical time scale
\citep{lev97}. 

In some astrophysical plasmas such as solar corona and stellar winds
thermal conduction can be important in heat transport.
We include the so-called classical Spitzer conduction for a fully ionized 
gas \citep{spi62}:

\begin{equation}
{S_{cl}=-k\nabla_r T},
\end{equation}
\begin{equation}
{k={1.84 \times 10^{-5}T^{5/2} \over \ln \Lambda_c}},
\end{equation}
where $\ln \Lambda_c = 29.7 + \ln n_e^{-1/2} (T/10^6K)$
and $n_e$ is the electron number density.
This classical formula is based on the assumption that the mean free path of 
the electrons is short compared to temperature scale height.
However, when temperature scale height is comparable or shorter than 
the electron mean free path, thermal conduction reaches 
the saturation limit and heat flux is no longer equal to $k \nabla_r T$.
In the saturation limit,
heat flux is given approximately by
\begin{equation}
S_{sat}\approx -0.4({{2k_BT} \over { \pi m_e}})^{1/2}~ n_e k_B T, 
\end{equation}
where $m_e$ is the electron mass \citep{cow77}. 
So we set $S = \min [S_{cl}, S_{sat}]$ as local heat flux.

With the classical formula the thermal conduction term  
becomes a diffusion term, $\overrightarrow \nabla_r \cdot (k \overrightarrow\nabla_r T)$. 
This type of partial differential equation can be solved {\it implicitly} 
by Cranck-Nicholson Scheme \citep{pres92}.
For the saturated formula, however, this cannot be applied.
So we solve the thermal conduction term {\it explicitly} and 
the integration time step is chosen as 
\begin{equation}
\Delta t^n = 0.1 \cdot \min ({ E \over {|\overrightarrow \nabla_r \cdot \overrightarrow S|}},
{ {\Delta r} \over {|u \pm c_s|}} ),
\end{equation}
where $c_s=(\gamma P/\rho)^{1/2}$ is the sound speed.

In grid-based Eulerian simulations a shock transition is typically
resolved over several grid zones with a strong temperature gradient, 
although the physical shock thickness may be much smaller than the size of 
one grid zone.
So un-physically rapid radiative cooling and heat transport can occur
across the numerical shock transition zones.
To avoid such numerical artifacts,
we identify zones that undergo shocking process and turn off 
radiative cooling and heat transport by thermal conduction inside
those zones.

We neglect magnetic fields and dust in our calculations.
If there exist dynamically significant, regular magnetic fields in the ambient 
medium, the assumed spherical symmetry would brake down.
Also coherent tangential magnetic fields would substantially reduce the radial
thermal conduction flux.
Dust may play some important roles through radiation pressure on dust and dust 
cooling in the cooled bubble shell in a more realistic situation. 
Proper treatments of such processes require radiative transfer, chemistry and
dust formation/destruction, which are beyond the scope of this study.
However, if the bubble shell stays photoionized at $10^4$ K by diffuse radiation, 
as we assume here, then the effects of dust are likely to be minimal.
\citet{smi01}, for example, included dust cooling in their simulations of 
multiple SNe and found that the dust cooling effects were only minor.

\section{Development of Wind Bubbles}
\subsection{Wind Models}

We first consider the development of a wind bubble to study the pre-supernova 
circumstellar environment. 
Massive early-type stars emit a steady stellar wind with constant terminal velocity 
$V_w$ and mass-loss rate $\dot {M_w} = dM_w/dt$.
We assume that the progenitor star has an initial mass $15 M_{\odot}$ and 
spends $\Delta t_w = 4 \times 10^6 yr$ in the main sequence phase.
The MS wind is characterized with a mass-loss rate, 
$\dot{M_w}=2.5 \times 10^{-7} \Myr$, 
and wind velocity, $V_w=2000 \kms$ \citep{smi06}. 
Here we consider only the MS wind, since the RSG wind produces a small shell with the 
radius of a few pc inside the bubble created by the MS wind.
The RSG wind shell would be swept up quickly by the SN ejecta during the early
free expansion stage, so interactions of the ejecta with this shell can be ignored.

We calculate the evolution of a bubble blown by collective stellar winds from
multiple massive stars in an association. 
The luminosity of the collective wind is the key physical parameter that determines 
the properties and evolution of the bubble
and it is given by 
\begin{equation}
 L_w =  {1 \over 2} \dot{M_w}V^2_w \cdot N_{star} = 
N_{star} \cdot (3.17 \times 10^{35} {\rm erg s^{-1}}). 
\end{equation}
We select the number of stars, $N_{star}=1, 5, 10, 30$ and 50,
because a typical OB association contains 10-100 early-type stars
(see Table 1).
In a realistic model, one needs to specify the initial mass function
of the association to obtain a total wind luminosity.
In our simplified model, however, $L_w$, is parameterized with $N_{star}$.
So a larger value of $L_w$ can be regarded as a higher wind luminosity 
either from a more massive star or from multiple stars.

We consider $n_{H,0}= 1 \cm3$ as the fiducial density of the ambient
ISM (WF models), 
where $\rho_0 = (2.34\times 10^{-24}{\rm g})n_{H,0}$.
Two additional values are considered: 
$n_{H,0}= 0.3\cm3$ (WL01 model) and $3 \cm3$ (WH01 model).
The ambient medium is assumed to be photoionized at $T=10^4$K.
So its sound speed is $c_s=14\kms$ and 
the pressure is $P_{ISM} =2.3\times 10^{-12} {\rm erg ~cm^{-3}}$.
With WZ01 and WZ10 models, we consider metal poor cases with $0.1\Zsun$.

The wind is realized numerically by injecting the gas flow with $V_w$ at 
the inner boundary ($r_w$) of the spherical simulation grid. 
The wind density at the inner boundary is given by  
$ \rho_w = N_{star} \cdot [\dot{M_w} / (4 \pi V_w r^2_w )]$. 
We use $N_r=$ 3000 uniform spherical grid zones in 150 pc radius with a spatial 
grid spacing of $\Delta r= 0.05$ pc.
Numerical convergence is often an important issue for hydrodynamic simulations 
with radiative cooling, because 
the cooling rate has a strong density dependence (\ie $\Lambda \propto \rho^2$).
So in order to test numerical convergence of the simulations
WF01 model is calculated with several different grid spacings. 
Fig. 1 shows the structure of the bubble in the simulations with 3 different
grid spacings. The contact discontinuity (CD) between the shocked wind and
the shocked ISM is around 23 pc, 
while the location of the conduction front (CF) is different  
in the three simulations.
The size of the hot cavity ($\sim$ 20pc) is smaller than
the so-called Field length 
$\lambda_F=\left[kT/(n_H^2L)\right ]^{1/2} \approx 56$ pc,
where $T\sim10^6$K and $n_H \sim 0.01\cm3$ are used
\citep{beg90}. 
So heat transport by thermal conduction is dominant over radiative cooling
inside the cavity. 
Heat transport from the hot shocked wind gas to the warm shocked
ISM reduces the temperature of the shocked wind gas, triggering radiative
cooling and compression.
These processes are not followed accurately enough to be converged
in the simulations, so the density and temperature distributions 
are significantly different inside the cavity 
among these simulations. 
However, the pressure distribution is very similar in the three calculations
and so the dynamical evolution of the wind bubble shell seems
to be converged.   
Radial positions of the CD and outer compression wave
seem to be converged within a few percent,
while the total energy inside the simulation volume seems to be
converged within $\sim$ 10 \%.
This fast convergence is achieved, 
because the density compression factor at the shell is not very large 
in the simulations ($\rho_{shell}/\rho_0 \lsim 10$).  
We also find that the differences in the density and temperature structures 
inside the cavity among the simulations with different grid spacings do not
affect significantly the evolution of SN explosions that will be
described in \S 4.2.

\subsection{Evolution of Wind Bubbles}

A typical bubble has four distinct regions from the central star outward: 1) freely
expanding wind with the density profile, $\rho\propto r^{-2}$, 2) a wind termination
shock facing inward and the shocked wind region, 3) a contact discontinuity (CD) between the
shocked wind region and the shocked ISM region, 
4) a forward shock propagating into the ambient ISM \citep{wea77}.
While the termination shock is strong, 
a compression wave may form, 
instead of the outer forward shock, 
depending on the ratio of wind ram pressure to the ambient pressure. 

Fig. 2 shows the gas density and temperature profiles at $t=10^6$ years  
in the models listed in Table 1.
Although the outermost structure is initially developed as a weak shock  
for all models, 
it becomes a compression wave with the propagation speed, $v \lsim c_s$
in some models, as the bubble expands out into the ambient medium.
For WF01 model, for example, the pressure increases only
by a factor of 1.5 across the compression wave.
Even for WF50 model the outer shock speeds is only $v_s \sim 20 \kms$.
Also the shock transition is much wider than that typically
found in adiabatic simulations using the same TVD code. 
In WF50 model the inner termination shock is at 10 pc,
the conduction front at $\sim 25$ pc, the CD at 41 pc, 
and the outer shock wave at $R_{bs} \sim 45$ pc.
Without thermal conduction, the hot cavity inside the CD would be 
almost isothermal with $T\approx 10^{7.6-7.8}$ K. 
But heat transport by thermal conduction enhances radiative cooling 
near the conduction front, leading to a strong temperature
gradient inside the cavity. 

Near the CD, the gas density changes by more than two orders of magnitude, 
so the gas cools quickly down to $T_{min}=10^4$K and forms a shell.  
The thickness of the shell is rather thick ($\Delta R_{bs} \sim 0.3 R_{bs}$).
The continuous injection of wind energy seems to be radiated away
efficiently near the conduction front.
For a spherical shell with inner radius, $R_{bs,i} = \alpha R_{bs}$, outer
radius, $R_{bs}$, and uniform density,
the shell density can be derived from the condition that the shell mass is 
equal to the swept-up mass: 
\begin{equation} 
{n_{shell} \over n_{H,0}} \approx (1-\alpha^3)^{-1}.
\end{equation}
So the density enhancement factor is small for thick shells found in our
simulations.  For WF01 model, for example, $\alpha \approx 0.7$ and 
$n_{shell}/n_{H,0} \approx 1.5$.

From Fig. 2 we can see that the size of bubble increases 
with $N_{star}$ and decreases with $n_{H,0}$. 
\citet{wea77} showed that the bubble would evolve self-similarly 
and the outer radius of the shell would grow 
as $R_{bs} \propto L_w^{1/5} \rho_0^{-1/5} t^{3/5}$,
if radiative cooling and the ambient pressure were negligible.
They also derived a slightly modified time dependence of 
$R_{bs} \propto t^{0.58}$ by numerical integrations of simplified conservation
equations where an approximate cooling rate and an ambient pressure 
with $T_{ISM}=$8000K were adopted.  
In our simulations the ambient pressure is not negligible
and the outward moving front is not a shock, 
so the self-similarity seems to be broken. 
  
Fig. 3 shows the evolution of  
$R_{bs}$ and swept-up shell mass, $M_{bs}=  (4\pi/3)R_{bs}^3 \rho_0 $,   
for all the models in Table 1.
For our models with $L_w= N_{star} (3.17\times 10^{35} {\rm erg s^{-1}}) $
and $n_{H,0}=1\cm3$,
the radius and mass of wind-blown bubble shell can be approximated by
\begin{equation}
R_{bs} \approx (22 {\rm pc}) N_{star}^{0.18}({t \over {10^6{\rm yrs}}})^{0.4},  
\end{equation}
and 
\begin{equation}
M_{bs} \approx (1.6\times 10^3 \Msun) N_{star}^{0.54}  
({t \over {10^6{\rm yrs}}})^{1.2}.  
\end{equation}
These relations are slightly different from what \citet{wea77} derived
through simplified numerical integrations described above.

Right panels of Fig. 3 show how $R_{bs}$ and $M_{bs}$ depend on $n_{H,0}$
or the metalicity $Z$.
These quantities do not scale with the ambient density in a simple way,
because the radiative energy loss depends not only on the gas density
but also on the detailed history of how heat is transported
from the hot cavity to the warm shell via thermal conduction. 
It seems that for $n_{H,0}<1 \cm3$ the swept-up mass is nearly 
independent of the ISM density, while $R_{bs} \propto n_{H,0}^{-1/3}$.
For high density medium ($n_{H,0}\gg1 \cm3$), the cooling time scale becomes
much shorter than the dynamical time, and so the numerical simulation
becomes increasingly difficult. 
For WH01 model with $n_{H,0}= 3 \cm3$, $M_{bs}$ 
is about 3 times that of WF01 models.
However, this trend cannot be generalized to higher density cases
because of nonlinear behavior of heat transport and radiative cooling.
For WZ01 with lower metalicity ($0.1 \Zsun$) the cooling is less efficient and so
the results are similar to those of WL01 model with lower ambient density.  

Bottom panels of Fig. 3 show the fraction of energy lost due to 
radiative cooling. 
The radiation energy is calculated as the energy lost from the simulation
volume, 
$E_{rad} (t) = L_w\cdot t + E_{tot,i} - E_{tot}(t)$,   
where $E_{tot}=E_{k}+E_{th}$ is the total energy inside the simulation
volume and $E_{tot,i}$ is its initial value. 
Here the thermal energy $E_{th}$ excludes the initial thermal energy of the ambient gas.
The energy loss due to radiative cooling is significant: about 60 \%
for WL01 and WLZ01 models and about 60-80 \% for WH01 and WF01-WF50 models.
For WF models the fraction of $E_{rad}$ is reduced somewhat for higher 
$L_w$ (larger $N_{star}$), because higher power wind creates hotter
cavities. 

\section{Multiple Supernova Explosion}

\subsection{Multiple Supernova Models}

Next we calculate the evolution of multiple supernovae exploded inside
the preexisting wind bubble created by the MS winds. 
So we take the final structures of the wind bubble simulations 
at $t=4 \times 10^6$yrs, described in the previous section, 
as the states of the circumstellar medium
prior to SN explosions. 

Each SN explosion is characterized by the explosion energy, $E_{ej}=10^{51}$erg, 
and the ejecta mass, $M_{ej}=10 \Msun$.
Then the ejecta can be approximated by
a uniform density core ($\rho_1$) and an outer envelope with a steep
power-law density profile ($\rho_2 \propto r^{-n}$),
expanding freely with the radial velocity, $v=r/t$ \citep{che89}. 
For a multiple explosion where $N_{SN}$ supernovae detonate simultaneously,
\ie total explosion energy $E_{SN}= N_{SN} E_{ej}$,
the density profile of the ejecta is defined by
\begin{equation}
{\rho_1 = Ft^{-3}~~ {\rm for} ~v \le v_t},
\end{equation}
\begin{equation}
{\rho_2 = Ft^{-3} \Biggl({{v} \over {v_t}}\Biggl)^{-n} ~~ {\rm for}~ v > v_t},
\end{equation}
where
\begin{equation}
{F={{N_{SN}} \over {4 \pi n}}{{[{3(n-3)M_{ej}]}^{5/2}} \over {{[10(n-5)E_{ej}]}^{3/2}}}},
\end{equation}
and 
\begin{equation}
{v_t = {\Biggl[{10 \over 3}{{(n-5)} \over {(n-3)}}{{E_{ej}} \over {{M}_{ej}}}\Biggr]}^{1/2}}.
\end{equation}
We adopt $n=10$, $v_t=3.16 \times 10^3 \kms$ and 
$F=N_{SN}\cdot(8.14 \times 10^7 {\rm g~ cm^{-3}~s^3})$.
We note that overall results of our calculations do not depend sensitively on
the specific values of $M_{ej}$ and $E_{ej}$ adopted.
As shown below, the evolution of a superbubble is controlled mainly by
the total explosion energy $E_{SN}$ rather than $E_{ej}$.
Also a SNR should lose its memory of ejecta mass after the swept-up mass becomes much 
greater than $M_{ej}$.

Table 2 shows the multiple SN explosion models considered here.
The number of SNe is given by $N_{SN1}$ and varies from 1 to $N_{star}$,
since $N_{star}$ is the total number of possible SNe 
inside a given bubble. 
As in the wind bubble stage, we consider three values of the ISM density:
$n_{H,0}= 1 \cm3$ as a fiducial value (SNF models), $0.3 \cm3$
(SNL01) and  $3 \cm3$ (SNH01).
Since SN explosions can occur sequentially with a time interval,
we also consider SNF02a model in which a second SN detonates 
$10^5$ years after the first SN.
This will be compared with SNF02b model in which two SNe explode simultaneously.
Again SNZ01 and SNZ10 models have a lower metalicity, $0.1\Zsun$.

We use the same grid setup as the wind simulations: \ie $N_r=$ 3000 uniform grid zones in 150 pc radius.
Again two additional simulations are carried out for SNF01 model 
with smaller grid spacings ($N_r=$ 6000 and 12000) 
to test numerical convergence. 
Radial position of the outer shock is converged within 0.5 \%,
while the total energy is converged within 5 \%.
The simulations are terminated at $t_f=10^6$ years.

\subsection{Evolution of Supernovae Remnants}

Fig. 4 shows the radial profile of the remnant for the single SN case 
(SNF01 model).
Initially, the SN profile described by Eqs. (14)-(17) is
superimposed onto the final output of the corresponding wind bubble model (WF01):
the wind termination shock is at 3 pc 
and the hot cavity extends out to 25 pc (solid lines in Fig. 4).
The compression wave has weakened and the shell thickness is large 
($\alpha \approx 0.43$),
so the density of swept up shell, $n_{shell} = 1.1 \cm3$,    
is slightly higher than the ambient density (see Eq. [11]). 
The SN blast shock wave passes through the pressure barrier
produced by the wind termination shock and travels quickly through 
the hot cavity.
The forward shock first interacts with the bubble shell at $\sim10^4$ years, 
pushing out the inner shell boundary and generating two secondary waves, 
i.e., a transmitted shock propagating outward into the bubble shell 
and a weak reflected compression wave moving inward into the ejecta.
A reverse shock forms in the SN ejecta around $5\times 10^3$ years 
and is reflected at the center around $7.5\times 10^3$ years. 
Afterward it becomes a traveling compression wave,
because the shock heated gas inside the cavity is hot. 
When this wave runs into the shell at $\sim10^5$ years, 
the shell is heated for the second time. 
 
By the termination time ($10^6$ years)
the shell is pushed out by the SNR to $R_{ss}=$67 pc with inner radius 
$R_{ss,i}=$54 pc and 
the swept up shell gas has $T = 10^4$ K and $n_{shell} \approx 4 \cm3$,
not much higher than the ambient medium.
On the other hand, the cavity is filled with the hot gas of 
$T \approx 10^6$ K and $n_H \approx 0.01 \cm3$.  
The outer shock weakens with expansion and slows down to
$\sim c_s$.

Fig. 5 compares the two different models of double explosions,
\ie sequential (SNF02a) and simultaneous (SNF02b) explosion models.
In SNF02a model, the second SN explosion is loaded $10^5$ years after
the first SN explosion.
Three snapshots are shown:
1) At $6.0\times 10^4$ years the outer shock in SNF02b model 
has expanded further out than the shock in SNF02a model.
2) At $1.2\times 10^5$ years, a little after loading of the second SN in SNF02a model, 
the second SN shock expands out to 35 pc, 
still well inside the shell at 42-46 pc generated by the first SN. 
3) At $2.0\times 10^5$ years the SN-driven shell has expanded {\it slightly} 
further out ($\sim$50 pc) in SNF02b model than in SNF02a model.
However, the difference is small enough, so we say the bubble size
is similar for the two models.

The left panels of Fig. 6 also show the comparison of the two double 
explosion models. 
The outer shock radius and shell mass of the two models are very similar. 
Here $E_{th}$ and $E_{k}$ are total thermal and kinetic energy
inside the simulation volume, respectively,
excluding the initial thermal energy of the ambient gas.
The time evolution of $E_{th}$ and $E_k$ shows that 
the initial kinetic energy of the SN ejecta is transferred to 
the thermal energy gradually in the beginning as the blast wave expands, 
but rather abruptly at $\sim 10^4$ years when the blast wave runs 
into the preexisting wind-blown shell. 
The thermalization process happens again at later times in SNF02a model,
when the second SN detonates at $10^5$ years and when its blast wave 
arrives at the first SN shell at $\sim 10^{5.2}$ years.
Interestingly, the latter event occurs at the similar time 
when the aforementioned compression wave hits the shell in SNF02b model, 
causing the second period of active radiative cooling.

Most of thermal energy is lost through radiative cooling.
Here the radiation energy is calculated by $E_{rad}(t) = E_{tot,i}-E_{tot}(t)$.
We note that the initial total energy, $E_{tot,i}$, 
includes both the SN energy and the energy of the preexisting wind bubble,
so $(E_{tot}+ E_{rad})/E_{SN}$ is greater than one.
At the termination time of our simulations, the final energy budget in the
two models are similar:  6\% in thermal energy, 7\% in kinetic energy, 
and 90\% lost through radiation energy.  
Figs. 5-6 imply that, for given total explosion 
energy, the different fashion of multiple explosion, 
\ie either sequential or simultaneous detonation, 
does not make significant differences in the size and mass of the final bubble
shell.

The right panels of Fig. 6 compare the models with 
different ambient density and different metalicity.
At lower density or with lower metal abundance, 
the bubble size is bigger and $E_{rad}$ is smaller, compared to higher
density models, 
but the shell mass is similar for all the models considered here. 

Simultaneous explosions with 1, 3, 5, and 10 SNe inside a wind bubble generated
by 10 massive stars are considered as SNF10a-d models.
Initial conditions are taken from WF10 model simulation:
the cavity is bounded by the CD at 56 pc and the outer compression wave is 
at 82 pc.
The upper two panels of Fig. 7 show the bubble structure 
at $10^6$ years for SNF10a-d models.
More energetic explosion pushes the shell further out and produces
a faster shock, leading to a larger and denser shell.
The lower panels of Fig. 7 show the profiles of SNRs in a uniform
ISM ($n_{H,0}=1 \cm3$) for comparison.
It shows that SN-driven bubbles are larger and hotter, if SNe explode 
inside a preexisting bubble. 
For example, the size of hot bubble for SNF10a-d models 
is 33-55 \% larger than that for SNe inside a uniform ISM, 
resulting in 2.4 - 3.7 times bigger volume of hot gas. 
Also the cavity temperature is 2-3 times higher in SNF10a-d models.
This implies that the volume filling factor and the temperature 
of the hot bubbles generated by core-collapse SNe 
would be underestimated by a factor of 2-4, 
if the preexisting bubble is not properly accounted for.

The left panels of Fig. 8 show the evolution of the bubble properties
for SNF10a-d models, while the right panels compare the 
SNZ10a-d models with lower metalicity. 
The radial position of the outermost shock increases 
as $R_{ss} \propto t^{0.6-0.8}$ before the arrival of the blast wave at the shell,
and afterward it increases very slowly, $R_{ss} \propto t^{0.1-0.15}$.
The size of SN-driven bubbles depends on both the SN explosion energy and
the size of pre-existing wind bubbles. 
According to the numerical results shown in Fig. 8,
$R_{ss}$ and $M_{ss}$ depends rather weakly on $E_{SN}$, 
\ie $R_{ss} \approx (85 {\rm pc}) N_{SN}^{0.1}$ and
$M_{ss} \approx (10^{4.8} \Msun)  N_{SN}^{0.3}$.

Again the evolution of $E_{th}$, $E_{k}$, and $E_{rad}$ indicates the 
complex thermalization history via shocks. 
The first major episode of radiative cooling occurs when the SN blast 
wave first hits the wind bubble at $t \sim 10^{4.5}$ years.  
When the compression wave generated by the reverse shock runs into the 
shell at $t \sim 10^{5.4}$ years, the shell is heated and compressed, 
leading to the second episode of rapid cooling.
In SNF10d model this behavior is most distinct. 
For SNF10a-d models, about 80-90 \% of $E_{SN}$ is lost via radiative cooling 
within $10^6$ years,
while only 10 \% remains as thermal energy and $\lsim$10 \% as kinetic energy. 
For SNZ10a-d models with lower metal abundance the radiative cooling is less
efficient compared to SNF10a-d models, so about 70 \% is lost, 
20-40 \% remains as thermal energy and 10 \% as kinetic energy.
Since the shell has a significant kinetic energy at that time, 
it would continue to expand slowly.
Exploration of later evolutionary stage requires more realistic treatments 
of radiative transfer, photoionization/heating, and non-equilibrium 
radiative cooling below $10^4$ K, which is beyond the scope of this study.

In the self-enrichment model of globular cluster formation, 
the shell of gas swept up by the blast wave from multiple SNe 
must cool and be confined gravitationally within the parent proto-globular 
cloud in order for the enrichment process to be effective
\citep{mor89,par99,par04}.
Thus the success of this model lies on how efficiently the SN explosion 
energy is radiated away to form gravitationally bound system.
According to Fig. 8 (SNZ10 models), 
in an average ISM with $n_{H,0}=1\cm3$ and 
$P_{ISM} =2.3\times 10^{-12} {\rm erg ~cm^{-3}}$, massive stars in an 
association with $N_{star} \approx 50$ would generate a bubble shell with 
$R_{ss} \approx 120$ pc and $ M_{ss} \approx 10^{5.5} \Msun$.
Incidentally this mass is similar to the characteristic mass of globular 
clusters. 
If 10 \% of explosion energy is left as the kinetic energy of the shell, 
then $E_k\approx 5 \times 10^{51} $ ergs.
With the gas temperature of $T=10^4$ K, the total thermal energy of the
photoionized shell is $E_{th} \approx 10^{51}$ ergs.  
But the gravitational energy, $|E_G|\sim G M_{ss}^2/R_{ss} \approx 10^{50}$ ergs, 
is too small to confine the shell gravitationally.
Thus the self-enrichment model would not work inside such low density 
clouds, but may operate in much higher density environment. 
If we blindly take the simple scaling relations shown in Figs. 6 and 8,
\ie $R_{ss} \propto n_{H,0}^{-1/3}$, but $M_{ss}\approx constant$, 
then the gravitational energy of the shell would increase with the ambient density
as  $|E_G| \propto n_{H,0}^{1/3}$.
In a much denser cloud higher ambient pressure would also help
the shell to be confined, possibly
resulting in a gravitationally bound system.

\subsection{Diffuse Radiation Emitted by Supernova Driven Bubbles}

As shown in the previous section, most of mechanical SN energy is 
transformed into thermal energy of the bubble shell and then  
lost via diffuse radiation. So radiative feedback is a significant
part of SN feedback into the surrounding ISM.
In order to study the photoionization/heating feedback of SN explosions
we calculate the diffuse radiation emitted by the bubble. 

First, we generate both continuum and line emissivities, 
$f_{\nu} = 4\pi j_{\nu}(T)/n_H^2$ (${\rm ergs~ cm^3 ~s^{-1} eV^{-1}}$) and 
$f_{line} = 4\pi j_{line}(T)/n_H^2$ (${\rm ergs~ cm^3 ~s^{-1}}$),
for $n_H=1 \cm3$ 
as a function of gas temperature for photon energy 
$E_{ph}({\rm eV})= h\nu=[E_{min}, E_{max}]=[1.36\times10^{-4},10^8]$,
using CLOUDY90 code (Ferland \etal 1998).
Collisional ionization equilibrium is assumed and no background
radiation field is adopted.
The emissivity tables are generated for two values of metalicity,
$\Zsun$ and $0.1\Zsun$.
Volume integrated spectrum of the diffuse radiation emitted by the bubble during
the integration time $t$ is calculated by 
\begin{equation}
 F_{\nu}(t) = \int_0^{t} dt \int_0^{R_{ss}(t)} 4\pi r^2 
\left[ f_{\nu} (T) + { f_{line}\over {\Delta E_{ph}}} \right] n_H^2(T, r) dr,
\end{equation}
where $F_{\nu}$ is given in units of ${\rm ergs~eV^{-1}}$ and
line emissivities divided by the width of a given photon energy 
bin are added to the corresponding continuum emissivity.
In practice, only $V(T,t)\equiv 4\pi \int dt \int n_H^2(T) r^2 dr$ is calculated
during simulations and then $F_{\nu}(t)= V(T,t) \cdot \left[ f_{\nu} (T) +  f_{line}/ 
\Delta E_{ph} \right]$ is calculated in a post-processing step.

The upper panel of Fig. 9 shows the evolution of $F_{\nu}(t)$ 
for a single SN model. 
Most of hard X-ray photons with $E_{ph}> 1$ keV are emitted early
before the blast wave runs into the wind bubble shell ($t<8\times 10^3$ years).
Afterward, the fluxes of lower energy photons 
increase with time, as the SN energy is deposited as heat 
into the bubble shell and radiated away mainly in UV and optical ranges.
The lower panel of Fig. 9 compares $F_{\nu}$ 
for SNF01 and SNZ01 models integrated to the final simulation time. 
In SNZ01 model metal lines are weaker, 
but H and He recombination continuum radiation is stronger, 
because the temperature is higher due to lower cooling rate. 

We also calculate the integrated energies and photon numbers of diffuse
radiation that can ionize hydrogen and helium atoms as follows:
\begin{equation}
\varepsilon_{E_i} = \int_{E_i}^{E_{max}} dE_{ph} F_{\nu}(t_f),
\end{equation}
\begin{equation}
\Phi_{E_i} = \int_{E_i}^{E_{max}} dE_{ph} { {F_{\nu}(t_f)}\over {E_{ph}} },
\end{equation}
where $E_i=13.6,$ 24.6, and 54.4 eV is the ionization energy for H, He, and 
He$^+$, respectively. 
These quantities are summarized in Table 3.
For SNF01 model, the total number of H ionizing photons is $\Phi_{13.6} 
\approx 10^{61}$ and the total energy carried away with those photons is
$\varepsilon_{13.6} \approx 0.55E_{SN}$. 
The total number of H ionizing photons scales roughly
with the SN explosion energy as $\Phi_{13.6} \sim  10^{61} N_{SN}$.

Diffuse radiation is enhanced overall in the models 
where a denser and hotter shell is produced. 
Comparing the results of SNL01, SNF01 and SNH01 models in Table 3, 
we can see that more diffuse photons are produced at denser environments. 
A SN with higher explosion energy generates a faster blast wave and 
produces a denser and hotter shell, emitting harder diffuse radiation. 
The results of SNF10a-d models show such trends. 
On the other hand, the comparison of SNF02a and SNF02b models indicate 
that the simultaneous double explosion generates more ionizing photons
than the sequential double explosion,
although the total radiation energy ($E_{rad}$) is similar in the two models.
SNZ models with lower metalicity produce less ionizing radiation
and more non-ionizing radiation,
compared to the models with solar metalicity.

Fig. 10 shows the evolution of $\varepsilon_{E_i}$ 
for SNF10a-d and SNZ10a-d models.
Most of ionizing photons are emitted during $10^{4.5}-10^5$ years when
the SN shock propagates into the bubble shell. 
For models with $N_{SN}=10$ ionizing photons are emitted actively
again when the reflected compression wave runs into the shell
around $10^{5.3}$ years (see also Fig. 8).  
 
\section{Summary}

We study multiple supernova (SN) explosions inside a preexisting cavity 
blown up by stellar winds from massive progenitor stars 
through numerical hydrodynamic simulations. 
Calculations are performed with a grid-based Eulerian code in one dimensional 
spherical symmetry, including radiative cooling and thermal conduction.
Heat transport by thermal conduction is important inside the hot cavity, 
while radiative cooling is dominant in the bubble shell.
We consider models with a wide range of physical parameters:
the ambient density, $0.3\le n_{H,0}\le 3 \cm3$, 
the metalicity, $\Zsun$ and $0.1\Zsun$,
the number of massive stars, $N_{star}=1-50$,
and the number of supernovae, $N_{SN}=1-10$.
In order to study the radiative feedback effects, we also
calculate the ionizing diffuse radiation emitted by the SN-driven bubble.

We first calculate the development of a bubble driven by
collective winds from $N_{star}$ massive stars with  
$L_w = N_{star}\cdot (3.17 \times 10^{35} {\rm erg s^{-1}})$.
Characteristic structures of wind bubbles are developed initially:
an inward facing termination shock, a conduction front near the contact 
discontinuity, and a forward expanding shock. 
The outermost shock is not strong and becomes a compression 
wave, because the wind ram pressure decreases as the bubble expands. 
At the end of the main sequence ($\sim 4\times 10^6$ years), 
the outer boundary of the bubble shell (\ie radius of the compression wave)
is $R_{bs} \approx (60 {\rm pc}) N_{star}^{0.18}$,
and the inner boundary is $R_{bs,i} \approx (0.7-0.9)R_{bs}$
for $n_{H,0}=1\cm3$.
So the shell is rather thick and the enhancement factor of
the shell density is small, \ie $n_{shell}/n_{H,0} =  1.5 - 3.7$. 
The mass of the bubble shell is 
$ M_{bs} \approx (2\times 10^4 \Msun) N_{star}^{0.54}$.

The evolution of a supernova remnant inside a wind bubble is much more complex 
than that in a uniform medium.
It depends on the structure and size of the preexisting
bubble as well as the total explosion energy, 
$E_{SN}=N_{SN}\cdot 10^{51}$ ergs.
The forward shock travels quickly through the cavity and runs into
the bubble shell, resulting in a transmitted forward shock.
On the other hand, the reverse shock in the ejecta is reflected 
at the center and becomes a compression wave that travels outward.
When this wave runs into the shell, the shell is heated for the
second time, followed by the second period of active radiative cooling. 
The wave bounces off the shell and moves inward with a much reduced amplitude. 

At the termination time of the simulations ($t_f=10^6$ years),
the outer shock radius is $R_{ss} \approx 67$ pc for a single SN inside 
a bubble blown by one massive star (SNF01 model),
while $R_{ss}\approx 85$ pc for a single SN inside a bubble blown 
by 10 massive stars (SNF10a model).
In comparison, the shock radius for a single SN in a uniform ISM of the same
density at the same time is about 60 pc.
The radii of the hot cavity and the bubble shell 
can be bigger up to 50 \% 
and the cavity temperature is higher by 2-3 times
for SN explosions in a wind bubble considered here,
compared to the SN explosions inside a uniform ISM.
For multiple SN explosions inside a bubble blown by 10 massive stars,
the size and mass of the shell increase rather weakly 
with the total explosion energy as
$R_{ss} \approx (85 {\rm pc}) N_{SN}^{0.1}$ and
a mass of $M_{ss} \approx (10^{4.8} \Msun)  N_{SN}^{0.3}$.
For $N_{SN}=50$, for example, the SN-driven shell would have typically 
$R_{ss} \sim 125$ pc and $M_{ss} \sim 10^{5.3} \Msun$ for 
the average ISM. 

Kinetic energy of the SN is transfered to thermal energy via various shocks.
About 20-60 \% of the explosion energy is radiated away after the SN shock 
hits the wind bubble shell, and then 20-40 \% is lost after the compression
wave crushes into the shell later. 
Although the final energy budget at the termination time of our simulations
depends on both the wind bubble structure and $E_{SN}$,
in most cases with solar metalicity, 
$E_{rad}/E_{SN} \approx 0.8-0.9$,  $E_{th}/E_{SN} \sim 0.1$ 
and $E_{k}/E_{SN} \lsim 0.1$. 
Ionizing photons of diffuse radiation account for 
up to $\varepsilon_{13.6}/E_{SN} \sim 0.55$,
while the total number of ionizing photons scales as 
$\Phi_{13.6}\sim 10^{61}N_{SN}$. 
For the cases with $0.1\Zsun$, the radiative loss is reduced 
a little bit, \ie $E_{rad}/E_{SN} \sim 0.7$, but
the fraction of ionizing radiation is reduced 
a factor of 2-3, compared to the models with solar metalicity.
So a larger fraction of energy is lost via non-ionizing radiation
at metal-poor environments.

We suggest physically correct prescriptions for SN feedback in numerical 
simulations of galaxy formation should reflect the natures described
in this study.  

\ack
The authors would like to thank J. Kim, D. Ryu and an anonymous referee
for helpful comments on the paper.
This work was supported by the Korea Research Foundation Grant
funded by Korea Government (MOEHRD, Basic Research Promotion Fund)
(R04-2006-000-100590)
and by KOSEF through the Astrophysical Research Center
for the Structure and Evolution of Cosmos (ARCSEC).
The work of HK was also supported in part by Korea
Foundation for International Cooperation of Science \& Technology
(KICOS) through the Cavendish-KAIST Research Cooperation Center.

\clearpage

\begin{table}
\begin{center}
{\bf Table 1.}~~ Wind Model Parameters \\
\begin{tabular}{c c c c c c} \hline\hline
  model & $n_{H,0}$ ($cm^{-3}$) & $N_{star}$ & $Z/Z_{\odot}$ &  \\  \hline
     WF01  & 1 & 1 &1 \\
     WF02  & 1 & 2 &1 \\
     WF05  & 1 & 5 &1 \\
     WF10  & 1 & 10 &1 \\
     WF30  & 1 & 30 &1 \\
     WF50  & 1 & 50 &1 \\
     WL01  & 0.3 & 1 &1 \\
     WH01  & 3 & 1   &1 \\
     WZ01  & 1 & 1  &0.1 \\
     WZ10  & 1 & 10 &0.1 \\
\hline
\end{tabular}
\end{center}
\end{table}


\begin{table*}[t]
\begin{center}
{\bf Table 2.}~~Multiple Supernova Explosion Models \\
\begin{tabular}{c c c c c c c} \hline\hline
  model & $n_{H,0}$ ($cm^{-3}$) & $N_{star}$ & $N_{SN1}$ & $N_{SN2}$ & $Z/Z_{\odot}$ & \\  \hline
   SNF01 & 1  & 1 & 1 & 0 & 1\\
   SNF02a & 1  & 2 & 1 & 1 & 1\\
   SNF02b & 1  & 2 & 2 & 0 & 1\\
   SNF10a & 1  & 10 & 1 & 0 & 1\\
   SNF10b & 1  & 10 & 3 & 0 & 1\\
   SNF10c & 1  & 10 & 5 & 0 & 1\\
   SNF10d & 1  & 10 & 10 & 0 & 1\\
   SNL01 & 0.3  & 1 & 1 & 0 & 1\\
   SNH01 & 3  & 1 & 1 & 0 & 1\\
   SNZ01 & 1  & 1 & 1 & 0 & 0.1\\
   SNZ10a & 1  & 10 & 1 & 0 & 0.1\\
   SNZ10b & 1  & 10 & 3 & 0 & 0.1\\
   SNZ10c & 1  & 10 & 5 & 0 & 0.1\\
   SNZ10d & 1  & 10 & 10 & 0 & 0.1\\
\hline
\end{tabular}
\end{center}
\end{table*}

\clearpage

\begin{table*}[t]
\begin{center}
{\bf Table 3.}~~Ionizing Photon Number and Energy of Diffuse Radiation \\
\begin{tabular}{c c c c c c c c} \hline\hline
model & log($\Phi_{13.6}$) & log($\Phi_{24.6}$)  & log($\Phi_{54.4}$)  
& $\varepsilon_{13.6}/E_{SN}$
 & $\varepsilon_{24.6}/E_{SN}$ 
& $\varepsilon_{54.4}/E_{SN}$
&\\  \hline
   SNF01 & 61.2 & 60.4 & 59.8 & 5.5(-1) & 1.6(-1) & 7.7(-2) \\
   SNF02a & 61.3 & 60.1 & 59.3 & 2.8(-1) & 4.1(-2) & 1.4(-2) \\
   SNF02b & 61.5 & 60.7 & 60.1 & 5.5(-1) & 1.9(-1) & 7.9(-2) \\
   SNF10a & 60.7 & 59.8 & 59.2 & 1.6(-1) & 4.5(-2) & 2.3(-2) \\
   SNF10b & 61.5 & 60.6 & 59.6 & 3.5(-1) & 7.4(-2) & 1.9(-2) \\
   SNF10c & 61.8 & 61.0 & 60.1 & 4.6(-1) & 1.2(-1) & 3.6(-2) \\
   SNF10d & 62.2 & 61.5 & 60.9 & 5.8(-1) & 2.2(-1) & 9.9(-2) \\
   SNL01 & 61.0 & 60.2 & 59.6 & 3.6(-1) & 1.2(-1) & 4.7(-2) \\
   SNH01 & 61.4 & 60.6 & 59.9 & 8.6(-1) & 2.9(-1) & 1.1(-1) \\
   SNZ01 & 60.8 & 60.1 & 59.2 & 2.2(-1) & 6.8(-2) & 1.9(-2) \\
   SNZ10a & 60.3 & 59.3 & 58.6 & 6.4(-2) & 1.3(-2) & 5.3(-3) \\
   SNZ10b & 61.2 & 60.5 & 59.3 & 1.7(-1) & 5.4(-2) & 8.8(-3) \\
   SNZ10c & 61.5 & 60.8 & 59.9 & 2.4(-1) & 7.2(-2) & 1.9(-2) \\
   SNZ10d & 61.8 & 61.1 & 60.3 & 2.3(-1) & 7.4(-2) & 2.5(-2) \\

\hline
\end{tabular}

\end{center}
\end{table*}

\clearpage

\begin{figure}
\includegraphics[scale=0.7]{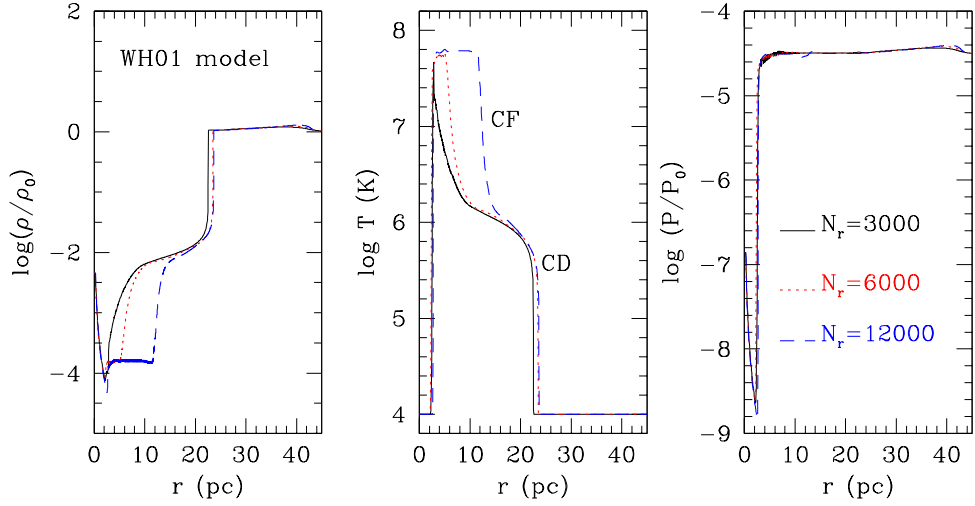}
\vskip -7.0cm
\caption{
Gas density, temperature, and pressure profiles at $t= 2.5\times10^6$ years
in WF01 model calculated with $N_r=$ 3000 (solid lines), 6000 (dotted), 
and 12000 (dashed) radial grid zones.
The density is given in units of $\rho_o=2.34\times 10^{-24} {\rm g~cm^{-3}}$,
and the pressure in units of $P_o = 9.36\times 10^{-8} {\rm erg~cm^{-3}}$.
}
\label{fig1}
\end{figure}

\clearpage
\begin{figure}
\includegraphics[scale=0.7]{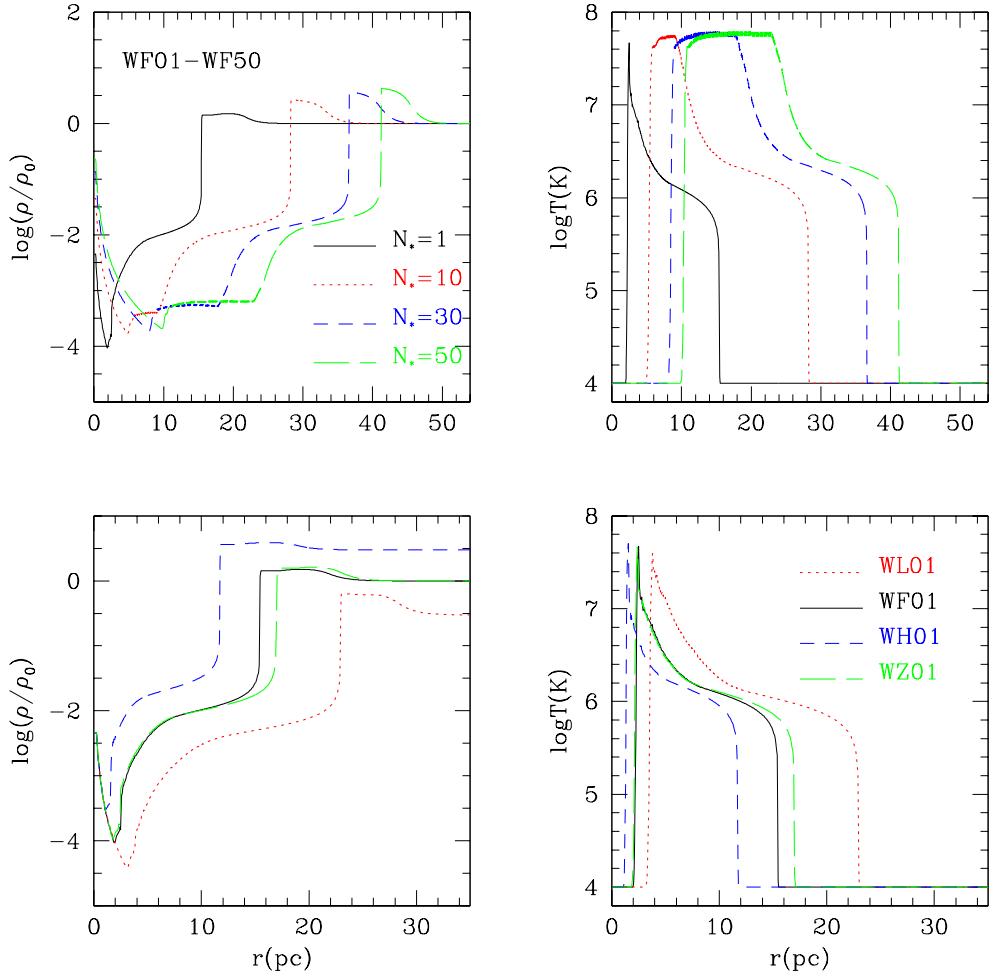}
\caption{
{\it Upper panels:} Gas density and temperature profiles at $t=10^6$ years
in WF01 (solid lines), WF10 (dotted), WF30 (dashed),
and WF50 (long dashed) models.
{\it Lower panels:} Gas density and temperature profile at $t=10^6$ yr
for WF01 (solid line), WL01 (dotted), WH01 (dashed), and WZ01 (long
dashed) models.
The density is given in units of $\rho_o=2.34\times 10^{-24} {\rm gcm^{-3}}$.
}
\label{fig2}
\end{figure}

\clearpage

\begin{figure}
\includegraphics[scale=0.7]{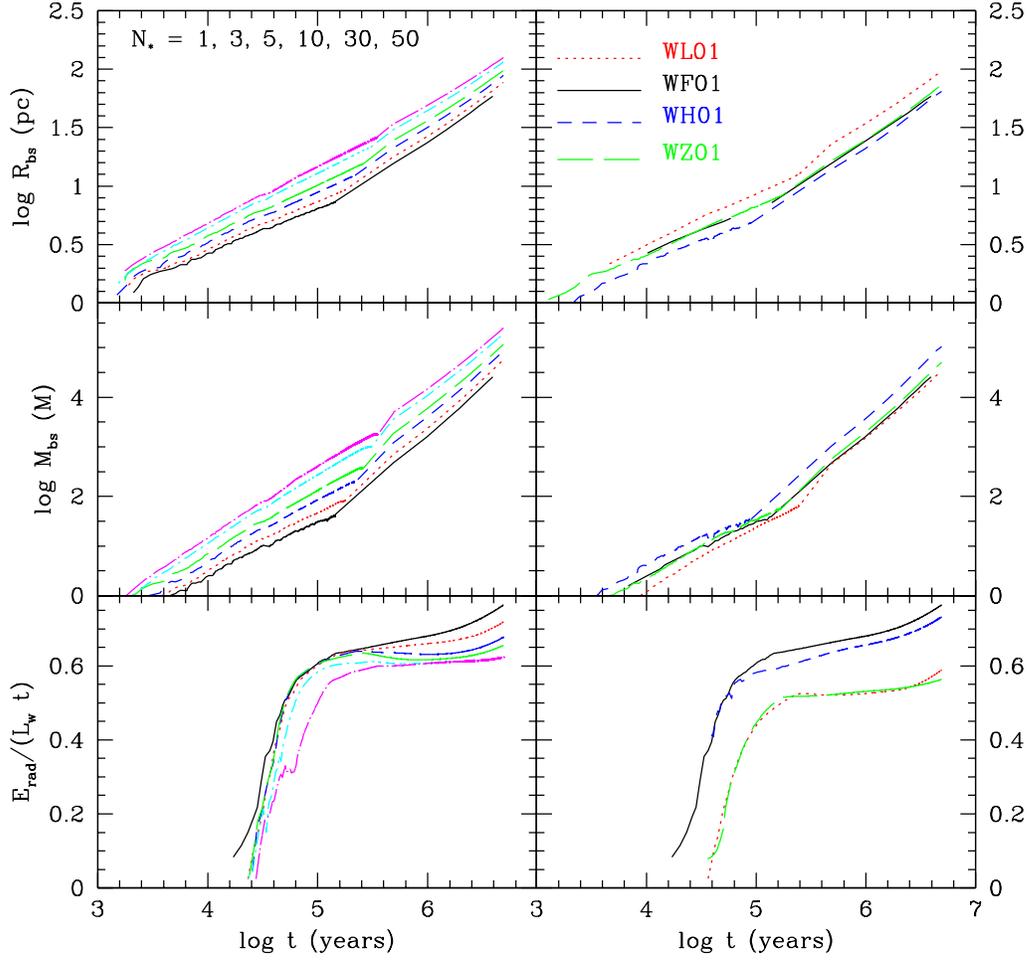}
\caption{
{\it Left panels:} Radial position of the outer compression wave (top),
swept-up mass (middle) of the wind-blown bubble shell, 
and the faction of energy lost due to radiative cooling
for WF01 (solid lines), WF03 (dotted), WF05 (dashed), WF10 (long dashed),
WF30 (dot-dashed), and WF50 (dot-long dashed) models.
{\it Right panels:} Same as the left panels
but for WF01 (solid lines), WL01 (dotted), WH01 (dashed), 
and WZ01 (long dashed)models.  
The radiation energy $E_{rad}$ is given in units of the total energy injected by winds.
}
\label{fig3}
\end{figure}

\clearpage

\begin{figure}
\includegraphics[scale=0.7]{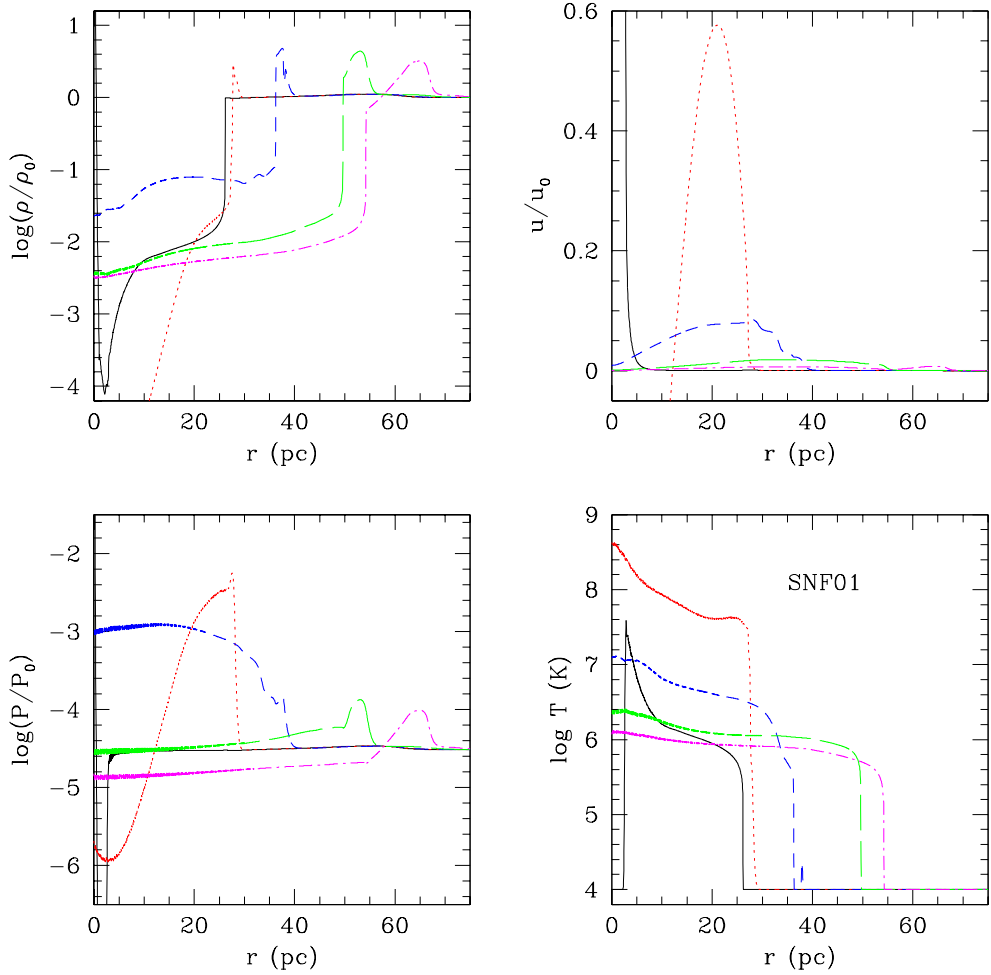}
\caption{
Time evolution of the supernova remnant from a single SN (SNF01 model)
at $t = 0$ (solid lines), $10^{4}$ years (dotted),
$10^{5}$ years (dashed),
$5. \times 10^{5}$ years (long dashed), and $ 10^{6}$ years
(dot-dashed).
The normalization constants are : $\rho_o=2.34 \times 10^{-24} {\rm gcm^{-3}}$,
$u_o=2000\kms$, and $P_o=9.37 \times 10^{-8}{\rm ergcm^{-3}}$.
}
\label{fig4}
\end{figure}

\clearpage

\begin{figure}
\includegraphics[scale=0.7]{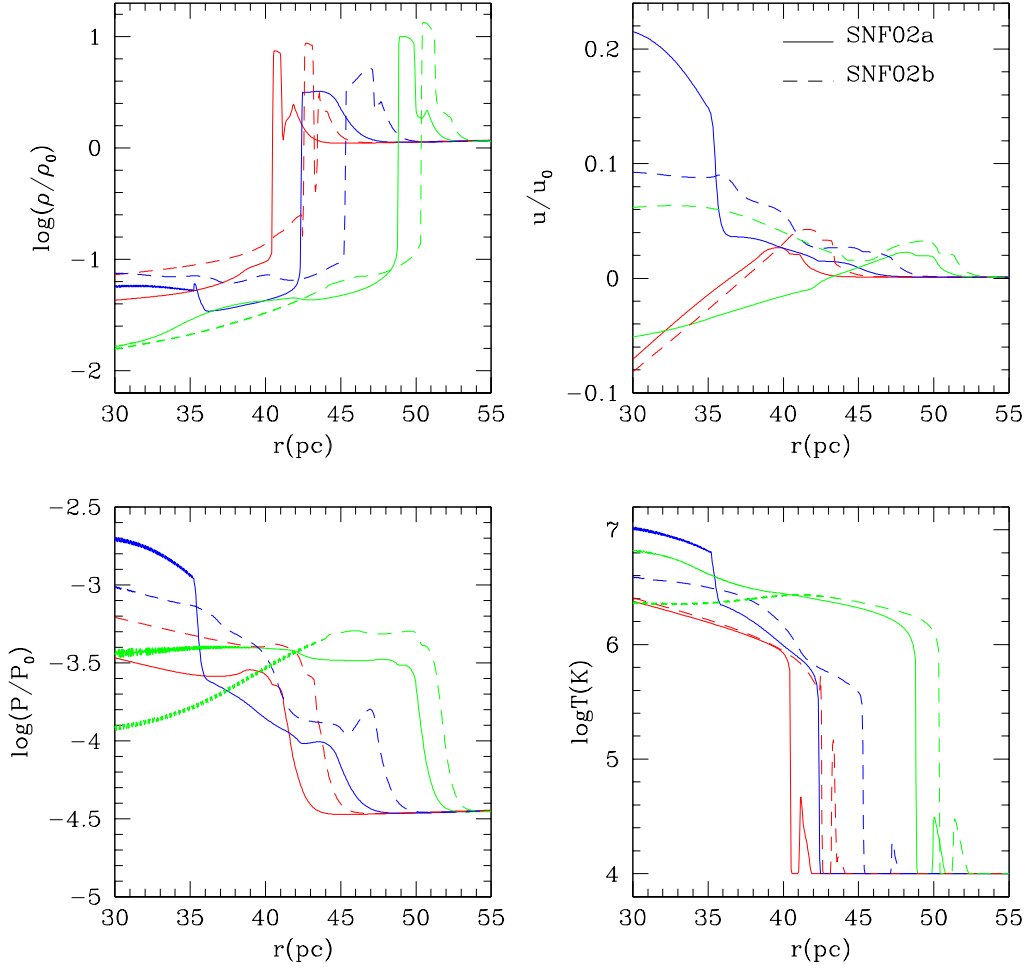}
\caption{
Time evolution of the supernova remnant for double SN explosion cases.
Solid lines are for the sequential explosion model (SNF02a),
while dashed lines are for the simultaneous explosion model (SNF02b).
Three profiles correspond to the remnants at $t=6 \times 10^4 yr$,
$1.2 \times 10^5 yr$, and $2 \times 10^5 yr$, from inside to outside,
respectively.
The normalization constants are the same as Fig. 3.
}
\label{fig5}
\end{figure}

\clearpage

\begin{figure}
\includegraphics[scale=0.7]{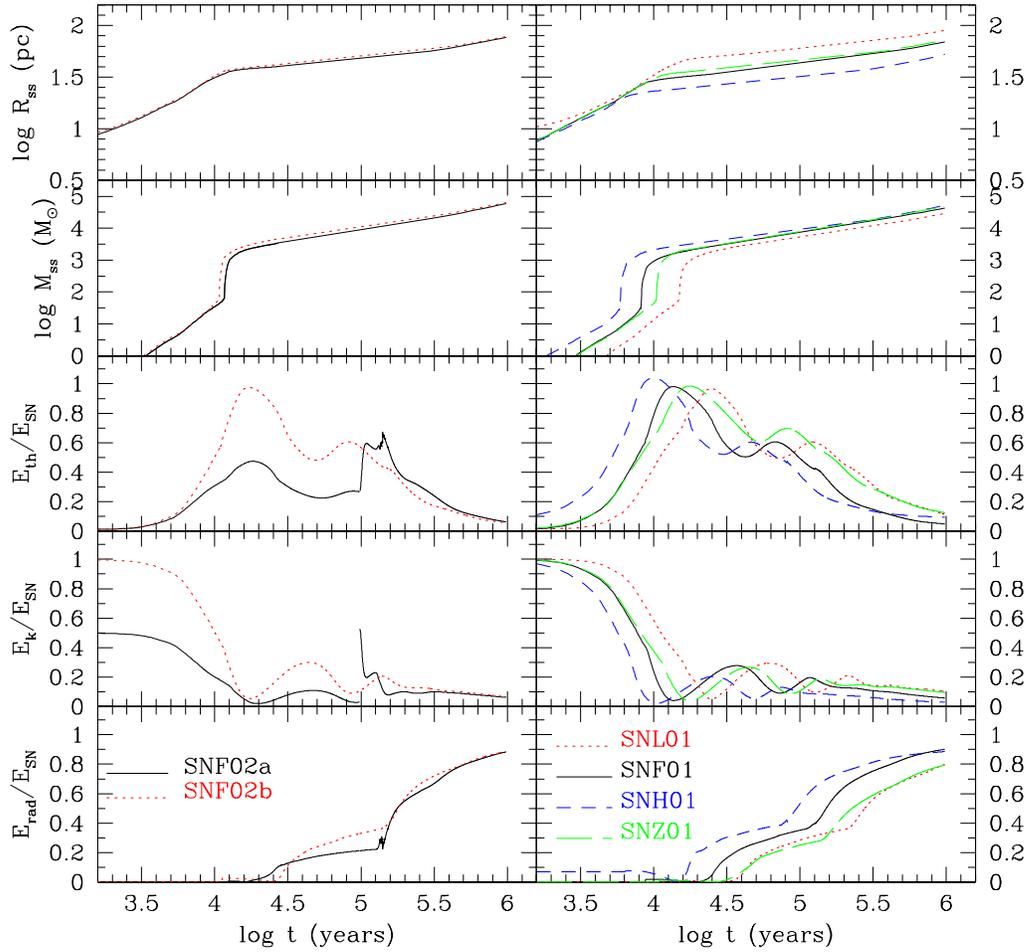}
\caption{
{\it Left panels}: Radial position of the outer shock,
swept-up mass of the SN-driven shell, thermal energy, kinetic energy,
and radiation energy (from top to bottom, respectively)
are shown for SNF02a (solid lines) and SNF02b (dashed lines) models.
{\it Right panels}: Same as the left panels except that the results for
are shown for SNF01 (solid lines), SNL01 (dotted lines), and
SNH01 (dashed), and SNZ01 (long dashed) models.
}
\label{fig6}
\end{figure}

\clearpage

\begin{figure}
\includegraphics[scale=0.7]{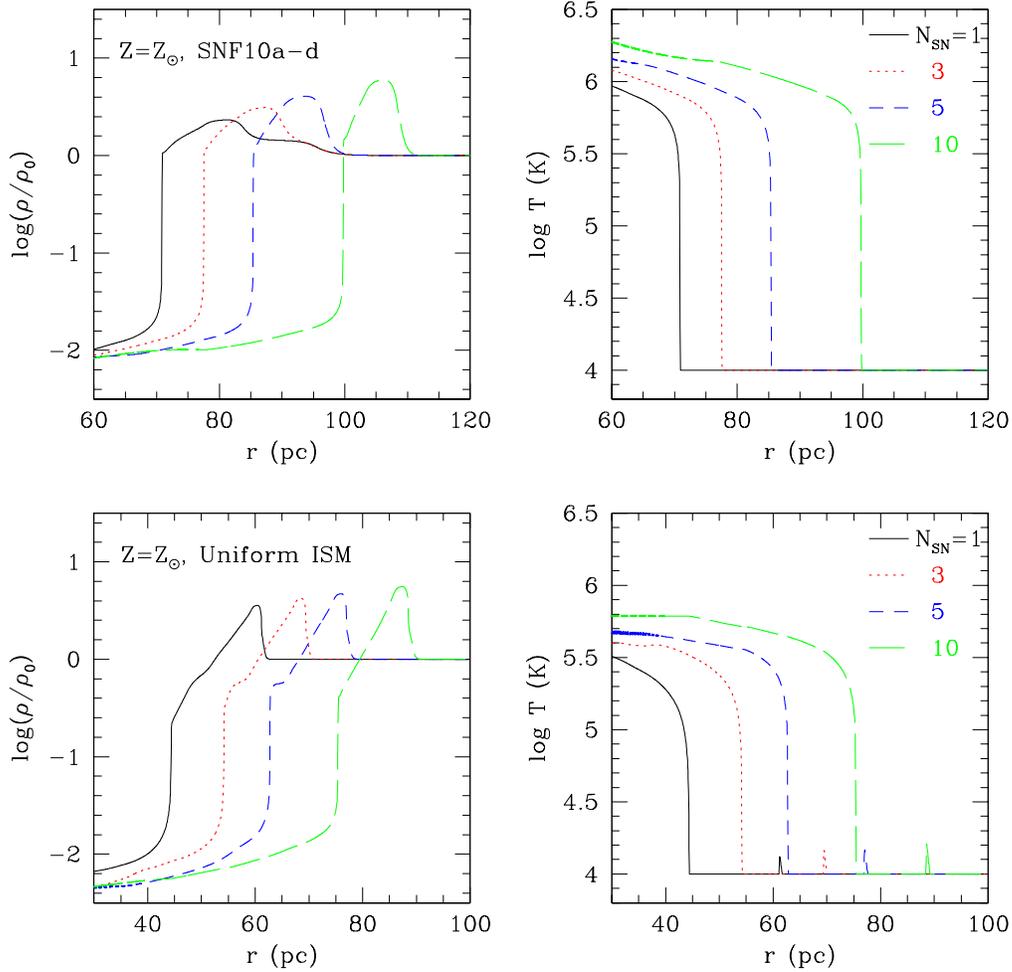}
\caption{
{\it Upper panels:} Supernova explosions inside a bubble created by 
10 massive stars (WF10 model).
SNF10a (solid lines), SNF10b (dotted),
SNF10c (dashed), and SNF10d (long dashed)
are shown at $t= 10^6 $years.
{\it Lower panels:} Supernova explosions inside a uniform ISM
of $n_{H,0}=1\cm3$. 
The explosion energy is $E_{SN}= N_{SN} 10^{51}$ ergs, where
$N_{SN}$ = 1 (solid lines), 3 (dotted), 5 (dashed), and 10 (long dashed).  
Here $\rho_o=2.34 \times 10^{-24} {\rm gcm^{-3}}$.
}
\label{fig7}
\end{figure}

\clearpage

\begin{figure}
\includegraphics[scale=0.7]{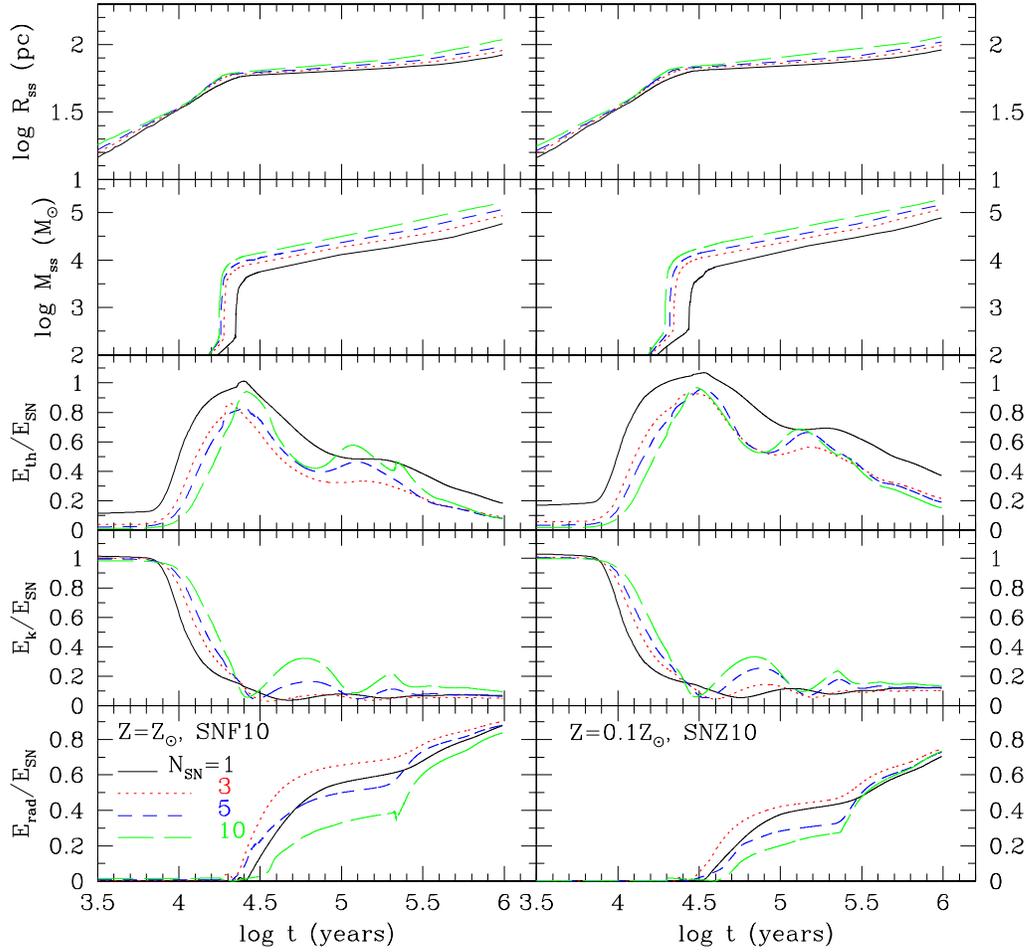}
\caption{
{\it Left panels}: Radial position of the outer shock wave,
swept-up mass of the SN-driven shell, thermal energy, kinetic energy,
and the energy lost due to radiative cooling (from top to bottom,
respectively) for SNF10a (solid lines), SNF10b (dotted), SNF10c (dashed),
and SNF10d (long dashed) are shown.
{\it Right panels}: Same as the left panels except that 
$Z=0.1Z_{\odot}$ is assumed.
}
\label{fig8}
\end{figure}

\clearpage

\begin{figure}
\includegraphics[scale=0.7]{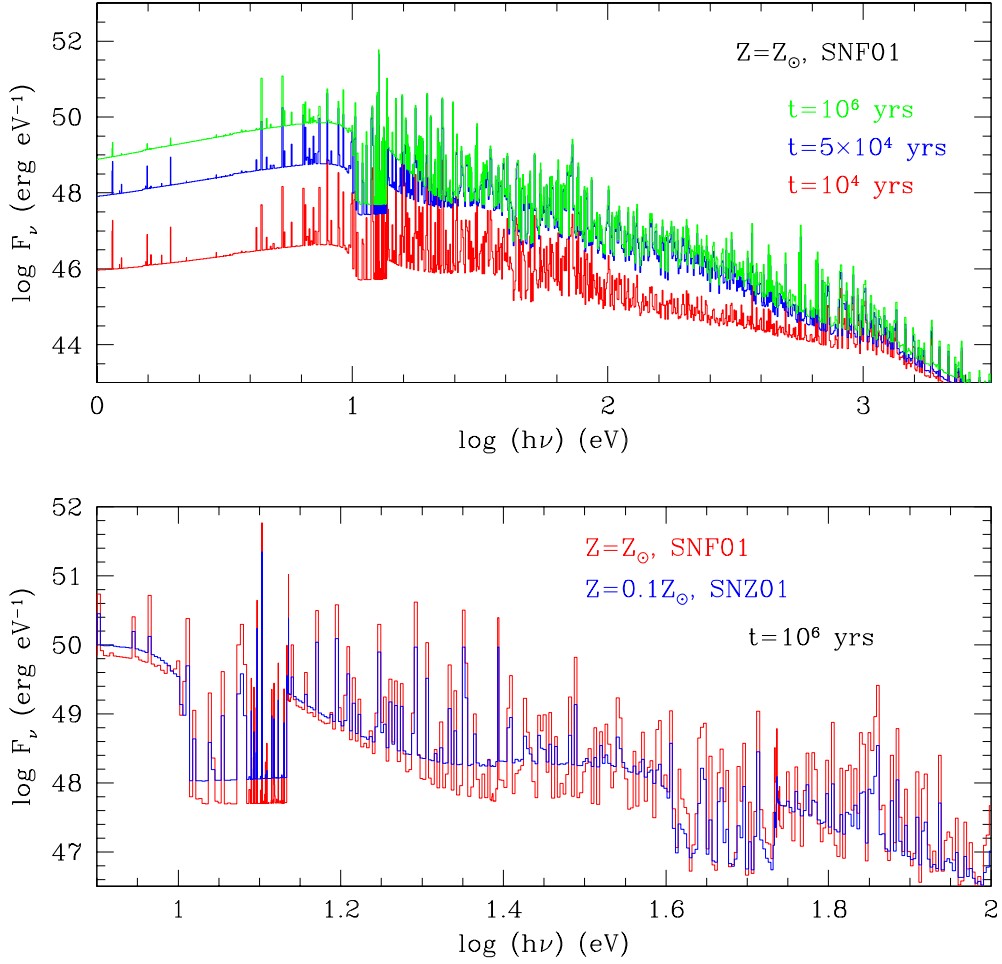}
\caption{
{\it Upper panel:} Integrated spectrum of diffuse radiation
emitted by the bubble for SNF01 model at $t = 10^4$ years (red),
$5\times 10^4$ years (blue), and $10^6$ years (green).
{\it Lower panel:} Integrated spectrum of diffuse radiation
emitted by the bubble in SNF01 model with $Z=Z_{\odot}$ (red) 
and SNZ01 with $Z=0.1Z_{\odot}$ (blue) models at $t=10^6$ years.
}
\label{fig9}
\end{figure}

\clearpage
\begin{figure}
\includegraphics[scale=0.7]{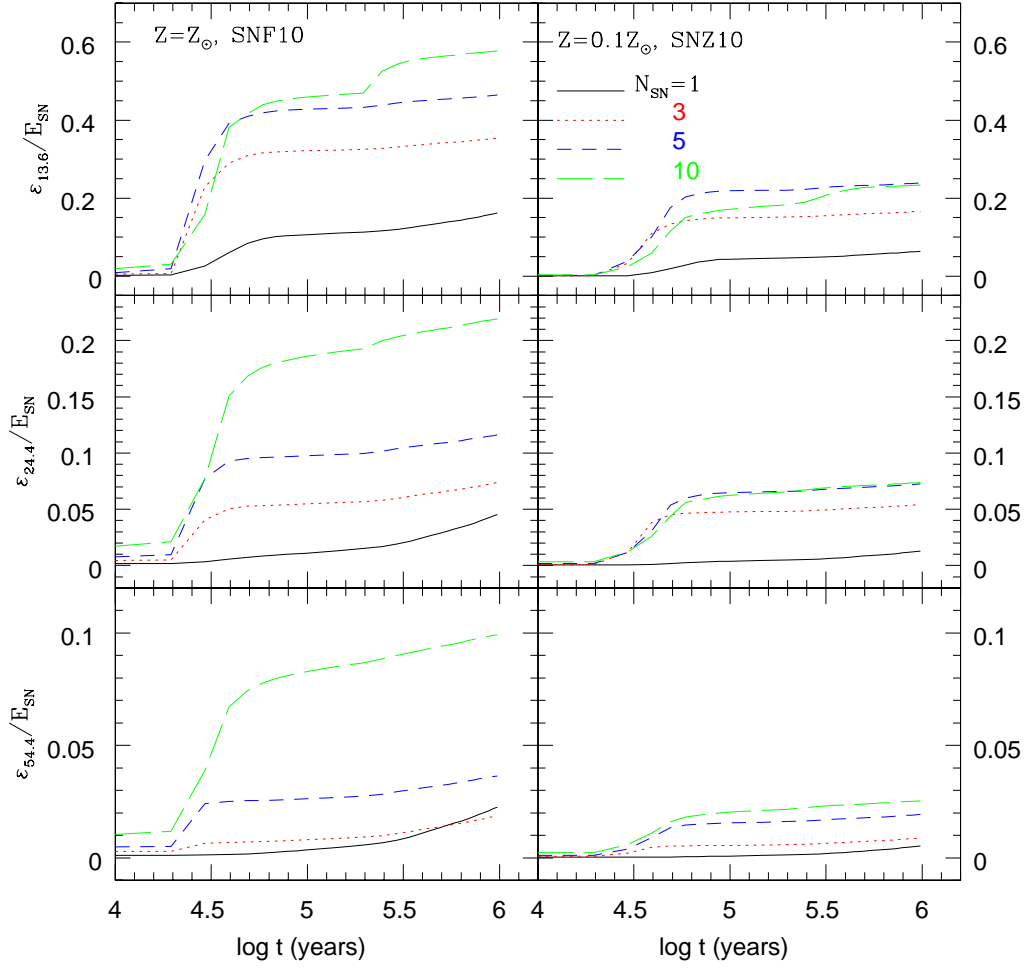}
\caption{ 
Time evolution of integrated ionizing energies of diffuse radiation
emitted by the bubble in SNF10 models with $Z=Z_{\odot}$
(left panels) 
and SNZ10 models with $Z=0.1Z_{\odot}$ (right panels).
The explosion energy is $E_{SN}= N_{SN} 10^{51}$ ergs, where
$N_{SN}$ = 1 (solid lines), 3 (dotted), 5 (dashed), and 10 (long dashed).  
Here $\rho_o=2.34 \times 10^{-24} {\rm gcm^{-3}}$.
}
\label{fig10}
\end{figure}

\clearpage

\end{document}